\documentclass[twocolumn,superscriptaddress,aps,prb]{revtex4-2}

\usepackage{graphicx}
\usepackage{dcolumn}
\usepackage{bm}
\usepackage{subfigure}
\usepackage{amsmath}
\usepackage{mathrsfs}
\usepackage{amssymb}
\usepackage{setspace}
\usepackage{array}
\usepackage{multirow}
\usepackage{float}
\usepackage{flushend}
\usepackage{footmisc}

\usepackage[pdfstartview=FitH,
CJKbookmarks=true,
colorlinks,
linkcolor=blue,
anchorcolor=blue,
citecolor=blue,
urlcolor=blue,
]{hyperref}

\usepackage{footmisc}

\renewcommand{\footnoterule}{
	\kern -4pt  
	\hrule width 0.18\linewidth height 0.6pt
	\kern 12pt 
}

\usepackage{footmisc}

\usepackage{braket}

\begin{document}

\preprint{APS/123-QED}

\title{Dissipation induced localization-delocalization transition in a flat band}
\author{Mingdi Xu}
\affiliation{School of Physics, Nankai University, Tianjin 300071, China}
\author{Zijun Wei}
\affiliation{School of Physics, Nankai University, Tianjin 300071, China}

\author{Xiang-Ping Jiang}
\email{2015iopjxp@gmail.com}
\affiliation{School of Physics, Hangzhou Normal University, Hangzhou, Zhejiang 311121, China}

\author{Lei Pan}%
\email{panlei@nankai.edu.cn}
\affiliation{School of Physics, Nankai University, Tianjin 300071, China}
\begin{abstract}
The interplay between dissipation and localization in quantum systems has garnered significant attention due to its potential to manipulate transport properties and induce phase transitions. In this work, we explore the dissipation-induced extended-localized transition in a flat band model, where the system's asymptotic state can be controlled by tailored dissipative operators. By analyzing the steady-state density matrix and dissipative dynamics, we demonstrate that dissipation is able to drive the system to states dominated by either extended or localized modes, irrespective of the initial conditions. The control mechanism relies on the phase properties of the dissipative operators, which selectively favor specific eigenstates of the Hamiltonian. Our findings reveal that dissipation can be harnessed to induce transitions between extended and localized phases, offering a novel approach to manipulate quantum transport in flat band systems. This work not only deepens our understanding of dissipation-induced phenomena in flat band systems but also provides a new avenue for controlling quantum states in open systems.
\end{abstract}

\maketitle

\section{Introduction}

The phenomenon of localization in quantum systems has been a cornerstone of condensed matter physics since Anderson's seminal work \cite{Anderson1958} on disorder-induced localization. Localization emerges from the interference of quantum waves in disordered potentials, resulting in the suppression of diffusion and the formation of spatially confined states. While Anderson localization is well-characterized in closed quantum systems, the introduction of dissipation--arising from coupling to external environments-introduces additional complexity to the problem. The study of open quantum systems has a well-established theoretical foundation and has recently garnered renewed attention due to breakthroughs in the controlled manipulation of dissipation and Hamiltonian parameters~ \cite{Exp1,Exp2,Exp3,Exp4,Exp5,Exp6,Exp7,Exp8}. As a result, open quantum systems with interaction \cite{OpenMB1,OpenMB2,OpenMB3,OpenMB4,OpenMB5,OpenMB6,OpenMB7,OpenMB8,OpenMB9,OpenMB10,OpenMB11,OpenMB12,OpenMB13,OpenMB14,OpenMB15,OpenMB16,OpenMB17,OpenMB18,OpenMB19,OpenMB20,OpenMB21,OpenMB22,OpenMB23,OpenMB25,OpenMB26} and disorder or quasiperiodic potentials have become major topics of investigation \cite{OpenMBL1,OpenMBL2,OpenMBL3,OpenMBL4,OpenMBL5,OpenDisorder1,OpenDisorder2,OpenDisorder3,OpenDisorder4,OpenDisorder5,OpenDisorder6,OpenDisorder7,OpenDisorder8,OpenDisorder9,OpenDisorder10,OpenDisorder11,OpenDisorder12,OpenDisorder13,OpenDisorder14,OpenDisorder15,OpenDisorder16,OpenDisorder17,OpenDisorder18}. Conventionally, dissipative processes are regarded as detrimental to quantum coherence, thereby suppressing localization phenomena. Prior theoretical work has demonstrated that environment-induced dephasing can disrupt localization and facilitate enhanced quantum transport transport \cite{OpenDisorder7,OpenDisorder8,OpenDisorder9}. Consequently, the general view has been that localization is unstable when dissipation is present, since the coherence essential for localization is readily disrupted, resulting in steady states that are no longer localized.

However, recent advancements in the study of open quantum systems have demonstrated that, dissipation can play a constructive role in preserving or even inducing localized states \cite{Yusipov17}. Recent studies show that dissipation could be harnessed to induce a transition between delocalized and localized states in various quantum systems including disordered and quasi-periodic systems \cite{WYC_PRL,Jiang_3D}. Moreover, the use of tailored dissipative operators has also been shown to selectively favor specific eigenstates of the Hamiltonian, leading to the formation of non-thermal steady states such as many-body localization \cite{Yusipov18,WYC_MBL} and quantum many-body scars \cite{Diss_scar}. This control mechanism relies on the phase properties of the dissipative operators, which can be tuned to adjust the spatial phase structure of the system's eigenstates. 
%


The unique properties of flat-band localization (FBL) allow for the formation of localized states which provides an alternative pathway for inducing localization in quantum systems. Unlike disorder-induced localization relying on the presence of random potentials to disrupt translational symmetry, the FBL arises from the unique geometric and topological properties of the lattice structure which does not require disorder and instead leverages the precise engineering of the lattice's hopping parameters to create non-dispersive energy bands. As a result, FBL offers a highly controllable and disorder-free approach to achieving localized states, making it particularly attractive for studying localization phenomena in clean systems. Furthermore, the interplay between flat-band localization and other factors, such as interactions and dissipation, opens up new possibilities for exploring exotic phases and transitions in both theoretical and experimental settings. Hence, when dissipation is introduced into such systems, it expected that dissipation could significantly alter the system's dynamics and lead to novel phenomena, such as dissipation-induced transitions between extended and localized phases.

The study of dissipation in flat band systems is motivated by both fundamental and practical considerations. 
Recent experiments in ultracold atoms implemented localization in flat bands  \cite{BoYan_flatband,Flatband_exp} and the 1D  an experimental investigation of the transition from FBL to Anderson localization in a one-dimensional (1D) Tasaki model with synthetic lattice of ultracold \(^{87}\)Rb atoms was reported \cite{Flatband_exp}. Motivated by experimental results and theoretical advancements in open quantum systems, in this work, we study the impact of dissipation on a system based on the experimentally reported one-dimensional Tasaki lattice model. By introducing dissipative operators, we analyze the system's steady-state density matrix and dissipative dynamics, finding that the system can be driven into a steady-state that is dominated by either localized or extended states, irrespective of initial condition choice. By adjusting the phase parameter of the dissipative operators, dissipation-induced transition between localization and delocalization occurs. The control mechanism is robust against additional dephasing, highlighting the potential for practical applications in quantum simulation and transport manipulation. 


\begin{figure}
                \centering
                \includegraphics[width=\linewidth]{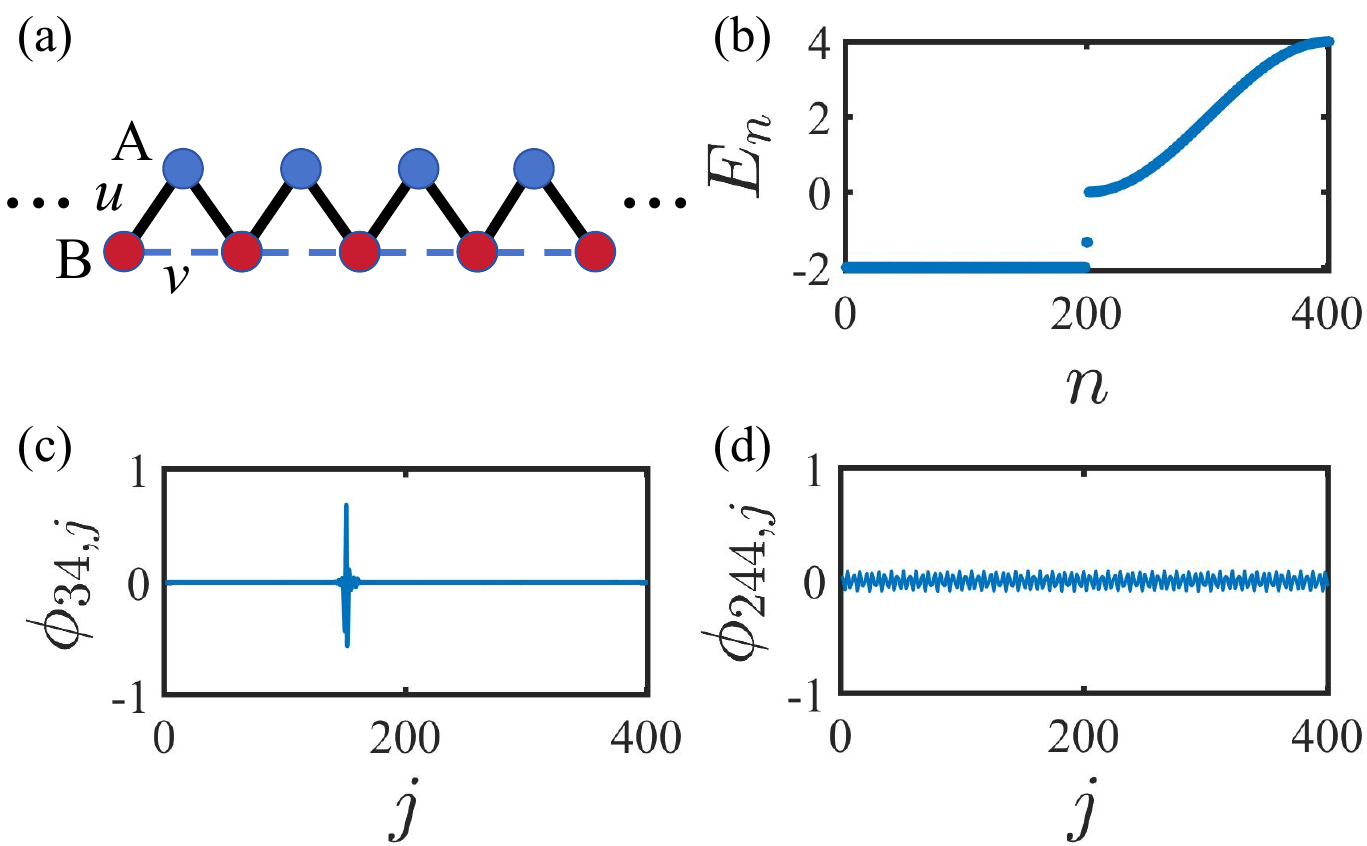}
                \caption{
                Properties of the energy spectrum and eigenstates of the one-dimensional Tasaki lattice.(a) The one-dimensional Tasaki lattice consists of two sublattices. The transition intensity $u$ is represented by a solid black line, while $v$ is represented by a dashed blue line. The blue lattice points represent the A sublattice, while the red lattice points represent the B sublattice.(b) The energy spectrum of the one-dimensional Tasaki lattice with open boundary conditions and $2L+1$ sites, where $L=200$, $u=\sqrt{2}$, $v=1$. Spatial distribution of the probability amplitude for the (c) 34th eigenstate and (d) 244th eigenstate.}
                \label{fig1}
            \end{figure}

\section{Model and flat band localization}

We consider the one-dimensional Tasaki lattice which consists of two distinct site types, denoted as \( A \) and \( B \). As depicted in Fig.\ref{fig1}(a), particles can transition between adjacent \( A \) and \( B \) sites, as well as between neighboring \( B \) sites. For a system comprising \( L \) unit cells, each containing one \( A \) and one \( B \) site, the total number of tunneling interactions between \( A \) and \( B \) sites is \( 2L-1 \), characterized by the coefficient \( u \). Additionally, there are \( L-1 \) interactions between \( B \) sites with a hopping rate \( v \). The Hamiltonian governing this system is expressed as

\begin{align}
H_{\text{Tasaki}} &= \sum_j \left( u\hat{c}_{j,A}^\dagger \hat{c}_{j,B} + u\hat{c}_{j,A}^\dagger \hat{c}_{j+1,B} + \text{H.c.} \right) \nonumber \\
&+ \sum_j \left( v\hat{c}_{j,B}^\dagger \hat{c}_{j+1,B} + \text{H.c.} \right), 
\label{Hamiltonian}
\end{align} 
  where \(\hat{c}_{j,A}^\dagger\) and \(\hat{c}_{j,A}\) ( \(\hat{c}_{j,B}^\dagger\) and \(\hat{c}_{j,B}\)) represent the creation and annihilation operators for particles on site $A$ ($B$) of the \( j \)th unit cell. The single-particle dispersion relation exhibits two bands: $E_\pm = |v| \cos k \pm \sqrt{|v|^2 \cos^2 k + 2|u|^2 (1 + \cos k)}.$ When the ratio satisfies \( r \equiv |u|/|v| = \sqrt{2} \), the lower band becomes entirely flat, with \( E_- = -2|v| \), as illustrated in Fig.\ref{fig1}(b). The eigenstates corresponding to this flat band are highly localized due to destructive interference in the hopping processes.

 The Fig.\ref{fig1}(b) plots the energy spectrum of the system, where the flat band and the two isolated points in the middle of spectrum correspond to localized eigenstates, which we refer to as the localized region. The remaining part of the spectrum, which follows a trigonometric function shape, corresponds to extended eigenstates, which we refer to as the extended region. In Fig.\ref{fig1}(c) and (d), we randomly select one eigenstate from the localized region and one from the extended region to illustrate their wavefunctions. 

We define $\hat{C}_{j}^\dagger$ and $\hat{C}_{j}$, $1\leq j \leq 2L + 1$, as a new set of creation and annihilation operators which satisfy the following relationship 
\begin{align}
	\hat{C}_j^\dagger = \hat{C}_{(j+1)/2, B}^\dagger, ~~~j\in \text{odd},\\
	\hat{C}_j^\dagger = \hat{C}_{j/2, A}^\dagger, ~~~j\in \text{even}
\end{align}
Then the Hamiltonian is rewritten as
\begin{equation}
	\begin{aligned}
		H_{\text{Tasaki}} = & \sum_{j=\text{even}} \left( u \hat{C}_{j}^\dagger \hat{C}_{j+1} + {H.c.} \right) \\
		& + \sum_{j=\text{odd}} \left( v \hat{C}_{j}^\dagger \hat{C}_{j+2} + H.c.\right).
	\end{aligned}
\end{equation}

\section{Theoretical Framework}
Suppose the system couples to a reservoir in which case the total Hamiltonian \(H_{T}\) of is given by 

\begin{align}
H_{T} = H_{S} + H_{R} + H_{SR},
\end{align}
where \(H_{S}\) and \(H_{R}\) represent the Hamiltonians of the system and reservoir, respectively, and \(H_{SR}\) describes the coupling between them. Under the Born-Markov approximation \cite{Moy1999,Breuer2002}, the dynamics of the system a is governed by the Lindblad master equation \cite{Lindblad1,Lindblad2} after tracing the reservoir's degrees 

\begin{align}
\frac{d\rho(t)}{dt} = \mathscr{L}[\rho(t)] = -i[H_{S}, \rho(t)] + \mathcal{D}[\rho(t)].
\end{align}
Here, \(\rho(t)\) is the density matrix of the system, and \(\mathscr{L}\) is the Liouvillian superoperator. The first term on the right-hand side represents the coherent evolution, while the second term \(\mathcal{D}[\rho(t)]\) captures the dissipative dynamics 

\begin{align}
\mathcal{D}[\rho(t)] = \sum_{j} \sum_{m=1}^{M} \Gamma^{(m)}_{j} \left( O^{(m)}_{j} \rho O^{(m)\dagger}_{j} - \frac{1}{2} \left\{ O^{(m)\dagger}_{j} O^{(m)}_{j}, \rho \right\} \right),
\end{align}
where $\{A, B\}\equiv AB+BA$ denotes the anticommutator, $O^{(m)}_{j}$ are the dissipation operator (jump operator), $j$ indexes the lattice sites, and $M$ represents the number of dissipation channels for each $j$ with corresponding strengths \(\Gamma^{(m)}_{j}\).

Using the Choi-Jamiolkowski isomorphism \cite{CJ1,CJ2}, the Lindblad equation can be reformulated in a vectorized form $\frac{d}{dt} |\rho\rangle = \mathscr{L} |\rho\rangle$,
where \(|\rho\rangle = \sum_{i,j} \rho_{i,j} |i\rangle \otimes |j\rangle\) is the vectorized form of the density matrix. Then the Liouvillian superoperator \(\mathscr{L}\) is expressed as \cite{Jiang_3D} 

	\begin{align}
	\mathscr{L}= & -i\left(H \otimes I-I \otimes H_{S}^{\mathrm{T}}\right) \nonumber \\
	& +\sum_{j}\sum_{m=1}^{M}\Bigg[2 O_{j}^{(m)} \otimes O^{*(m)}_{j}-O^{(m)\dagger}_{j} O_{j}^{(m)} \otimes I\nonumber \\
	&-I \otimes\left(O^{(m)\dagger}_{j} O_{j}^{(m)}\right)^{\mathrm{T}}\Bigg]. \label{Liouvillian}
\end{align}

The dynamics of an open quantum system is determined by the Liouvillian spectrum and the formal solution is given by \(|\rho(t)\rangle = e^{\mathscr{L}t} |\rho(0)\rangle\). 
The system eventually relaxes to a steady state defining by \(|\rho_{ss}\rangle = \lim_{t \to \infty} |\rho(t)\rangle\), which corresponds to the eigenstate of \(\mathscr{L}\) with zero eigenvalue. The behavior of the steady-state depends on the choice of jump operators in the Liouvillian superoperator.

\begin{figure}
	\centering
	\includegraphics[width=1\linewidth]{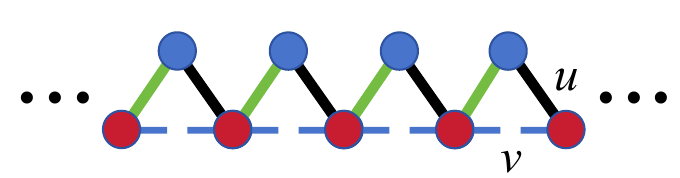}
	\caption{Schematic diagram of dissipation. The green solid line indicates the positions where the dissipation is applied.}
	\label{fig2}
\end{figure}

To explore the dissipation-induced localization, we consider dissipation operators as the following form 
\begin{equation}
	O_j= \left(\hat{C}_{j}^\dagger + e^{i\alpha} \hat{C}_{j+l}^\dagger\right)\left(\hat{C}_{j} - e^{i\alpha} \hat{C}_{j+l}\right),
\end{equation}
where the number of channel $M$ is set to $1$ and  \( j \) is odd. These operators initially appeared in earlier works~\cite{Jump1,Jump2}. Their realization employs a one-dimensional Bose-Hubbard system~\cite{BHchain}, while an alternative configuration using Raman optical lattices enables phases of $0$ or $\pi$~\cite{WYC_PRL}. 
Physically, each dissipation operator above acts on two sites ($j$ and $j+l$) and modifies the phase difference between them. For instance, when the dissipation phase is set $\alpha=0$,  the dissipation operator is expressed as $O_j= \left(\hat{C}_{j}^\dagger + \hat{C}_{j+l}^\dagger\right)\left(\hat{C}_{j} - \hat{C}_{j+l}\right)$, which synchronizes two sites from an anti-symmetric out-of-phase mode to a symmetric in-phase mode. Similarly, when the phase is $\alpha=\pi$,  the dissipation operator $O_j= \left(\hat{C}_{j}^\dagger -\hat{C}_{j+l}^\dagger\right)\left(\hat{C}_{j}+\hat{C}_{j+l}\right)$ annihilates  a symmetric in-phase mode from two sites and create an anti-symmetric out-of-phase one.  
 
The dissipation phase in jump operators critically influence the localization behavior of the system. Adjusting these phases allows selective population of spectral regions (low, middle, or high energy). This feature is essential for analyzing steady-state localization and nonequilibrium transitions, as discussed later.

\section{Dissipation-induced localization transition}

In this section, we will show that the dissipation operators introduced above can induce a localized or delocalized steady-state and the localization-delocalization transition occurs by tuning dissipation parameters. We first consider the case of $l=1$, with the schematic diagram of the dissipative effect shown in Fig.~\ref{fig2}, where the dissipation operator couples the $A$ and $B$ sites in a unit cell (green lines). To investigate the localization properties of the system under dissipation, we calculate distribution of the steady-state density matrix $\rho_{mn}=\langle \psi_m |\rho_{ss}|\psi_n\rangle$ in the eigenstate basis of the Hamiltonian, where $|\psi_m\rangle$ and $|\psi_n\rangle$ refer to  $m$-th and $n$-th eigenstates of the Hamiltonian, ordered by energy.

\begin{figure}
                \centering
                \includegraphics[width=1\linewidth]{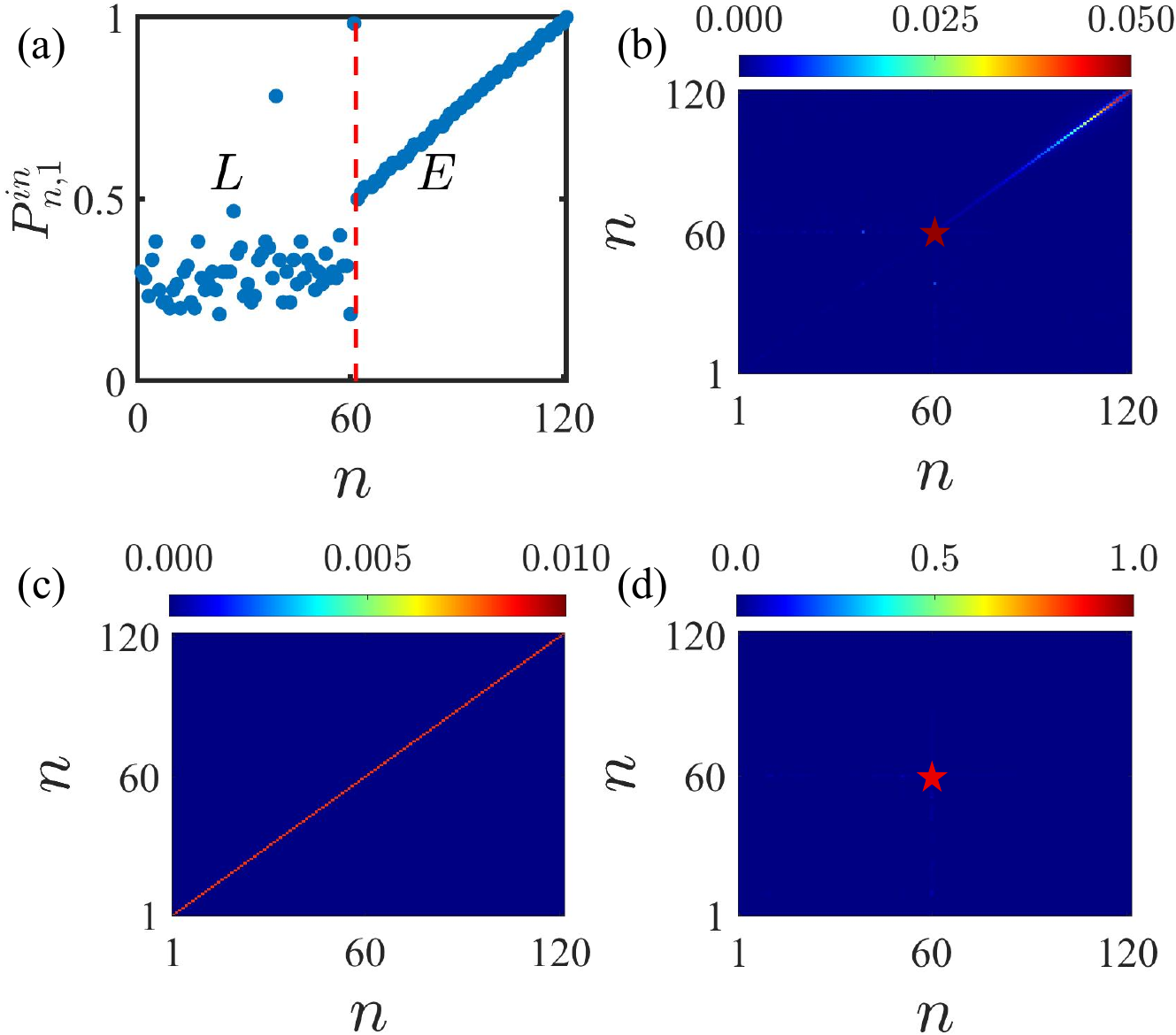}
                \caption{
                    Steady state induced by dissipation. 
                    (a) The proportion $P^{in}_{n,1}$ of in-phase site pairs among all dissipation-applied site pairs in each eigenstate. 
                    The localized states and extended states are located on the left and right sides of the red dashed line, respectively. 
                    Steady-state density matrix corresponding to the dissipation phase: (b) $\alpha=0$, (c) $\alpha=\pi/2$, (d) $\alpha=\pi$.The pentagram in (b) represents the 61st diagonal element of the density matrix, which has a value of 0.05.The pentagram in (d) represents the 60st diagonal element of the density matrix, which has a value of 0.89.}
                \label{fig3}
            \end{figure}
    
As shown in Fig.\ref{fig3}(b)-3(d), one can see that as the dissipative phase transitions from $0$ to \( \pi \), the steady-state distribution shifts from being predominantly composed of high-energy states (extended states) to being primarily composed of the localized eigenstate at the center of energy spectrum.

To understand how the steady-state configuration  connects the dissipative phase, we analyze the relative phase across adjacent lattice sites. For an eigenstate $\ket{\psi_n}$, we expand it as $ \left| \psi_n \right\rangle = \sum_{j=1}^{2L+1} \psi_{n,j} \hat{C}_j^\dagger \left| 0 \right\rangle$. The relative phase between the \( j \)-th lattice site and the \( (j+l) \)-th site is given by 
$\Delta\phi_{j,l}^n=\arg(\psi_{n,j})-\arg(\psi_{n,j+l}) $ where $\arg(\psi_{n,j})$ marks the argument of $\psi_{n,j}$. Then we calculate the number of lattice pairs affected by dissipation with a zero relative phase named by in-phase site pairs, denoted as \( N_{n,l}^{in} \), and compute the corresponding ratio $P_{n,l}^{in} = N_{n,l}^{in}/{N_t}$. Here \(N_{t}= L+1 - l \) is the total number of lattice site pairs related by dissipation.  Fig.~\ref{fig3}(a) plots the proportion of in-phase site pairs $P_{n,1}^{in}$ for each eigenstate. 

As illustrated in Fig.~\ref{fig3}(a), the distribution of $P_{n,1}^{in}$ among eigenstates indicates that extended states in the high-energy region possess higher values which increase monotonically with energy, while localized states in the low-energy region maintain relatively low and energy-insensitive values, with the exception of a particular localized eigenstate near the middle of the spectrum that exhibits an anomalously large $P_{n,l}^{in}$. That's to say, eigenstates on the lower-energy side of the spectrum exhibit a higher proportion of in-phase site pairs, whereas those at higher energies predominantly display out-of-phase configurations. This behavior stems from the consequence that the relative phase between any two nearest neighbor sites in an eigenstate is either $0$ (in phase) or $\pi$ (out of phase). Consequently, when the dissipation phase is set to $\alpha=0$, the system evolves toward a steady-state that is heavily weighted toward high-energy, spatially extended modes as plotted in  Fig.~\ref{fig3} (b), while when the dissipative phase is $\alpha=\pi$, the steady-state is  mainly composed of one of the eigenstates with the lowest value of \( P_{n,1}^{in} \) as shown in the  Fig.~\ref{fig3} (d).  We further  observe that, when the dissipative phase is $\alpha=\pi/2$, the steady-state distribution is uniform meaning that density matrix is proportional to identity matrix as shown in Fig.~\ref{fig3} (c). 
The emergence of this behavior stems from the Hermitian nature of the dissipation operator at $\alpha=\pi/2$, i. e., $O_{j}=\left(\hat{C}_{j}^\dagger + i\hat{C}_{j+l}^\dagger\right)\left(\hat{C}_{j} - i\hat{C}_{j+l}\right)=O_{j}^\dagger$, being Hermitian. Under this condition, dissipation drives the system toward complete thermalization, with the equilibrium density operator approaching a uniform distribution \cite{Longhi2023}. Consequently, this mechanism effectively suppresses all signatures of wavefunction localization.

\begin{figure}
        \centering
        \includegraphics[width=1\linewidth]{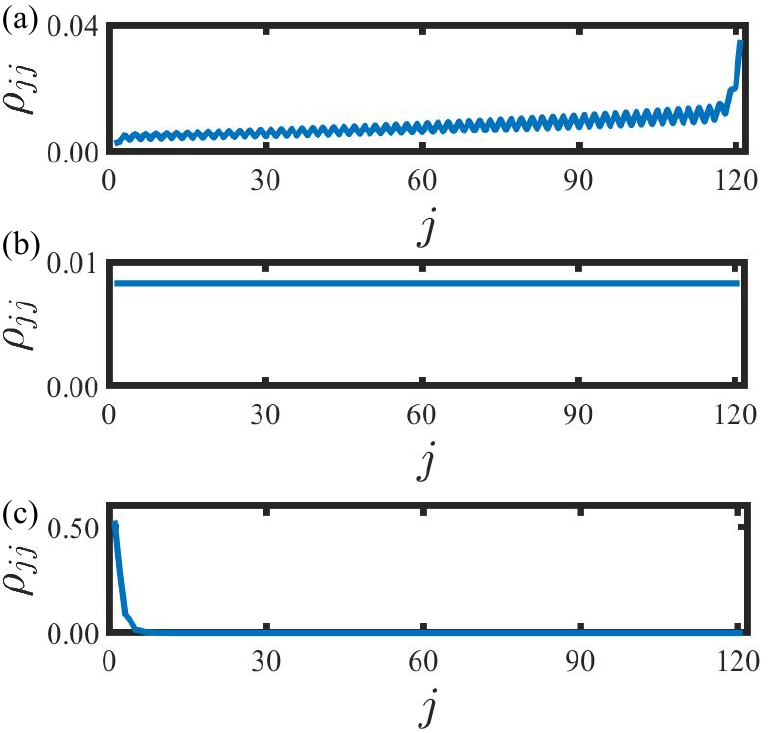}
        \caption{Distribution of the diagonal elements of the density matrix in the lattice basis. 
            The distributions correspond to: (a) $\alpha=0$, (b) $\alpha=\pi/2$, (c) $\alpha=\pi$.
        }
        \label{fig4}
    \end{figure}

To further characterize the localization feature of the steady-state, we calculate the spatial distribution  by transforming the steady-state density matrix into the real-space basis and then plot the its diagonal elements $\rho_{jj}$ as presented in Fig.~\ref{fig4}. The Fig.~\ref{fig4} (a) displays a delocalized distribution where the density is nonzero in the whole system when $\alpha = 0$, consistent with its eigenbasis distribution concentrating in high-energy extended states shown in the Fig.~\ref{fig3}(b). Conversely, steady-state at $\alpha = \pi$ shows localized spatial distribution as plotted Fig.~\ref{fig4} (a), since the steady-state density matrix in the eigenbasis primarily occupies localized modes. For the case of $\alpha = \pi/2$, the real-space density distribution is completely uniform as shown in the Fig.~\ref{fig4} (a) because the identity matrix is invariant  under any representation transformation.

Above results indicate that by tuning the dissipation phase $\alpha$, one can induce a transition between localized and extended (non-localized) steady states. 
To further see clearly the dissipation phase dependency to the steady-state, we further compute the diagonal elements corresponding to the localized modes which gives the proportion of localized state $P_l = \sum_{n=1}^{L+1} \rho_{nn}$. We find that as the dissipation phase \( \alpha \) increases, $P_l$ correspondingly grows meaning that this type of dissipation induces a transition from delocalization to localization in the steady-state.
We can also visualize the transition from perspective of dissipative dynamics. Since the steady-state is independent of initial states choice, if the initial state is prepared on a delocalized state, the system can be driven to a steady-state predominantly consisting of localized states, meaning that a localization transition is implemented by using a tailored dissipation $\alpha=\pi$, $l=1$. On the other hand, the localization is maintained if a localized initial state is prepared for this dissipation. 

From dynamic point of view, the localization information of a initial state can be preserved when the system relaxes to a localized steady-state.  Next, we introduce the quantum fidelity to portray this behavior. The quantum fidelity, which represents the overlap between initial
state $\rho_0$ and the time evolved state $\rho(t)$, defined as \cite{Fidelity1,Fidelity2}

\begin{equation}
	F[\rho(t), \rho_0] = \text{Tr}\left[\sqrt{\rho(t)^{1/2} \rho_0 \rho(t)^{1/2}}\right].
\end{equation}
Starting with a few representative initial states, we compute the dissipative dynamics and evaluate the quantum fidelity. 

\begin{figure}[!ht]
                \centering
                \includegraphics[width=1\linewidth]{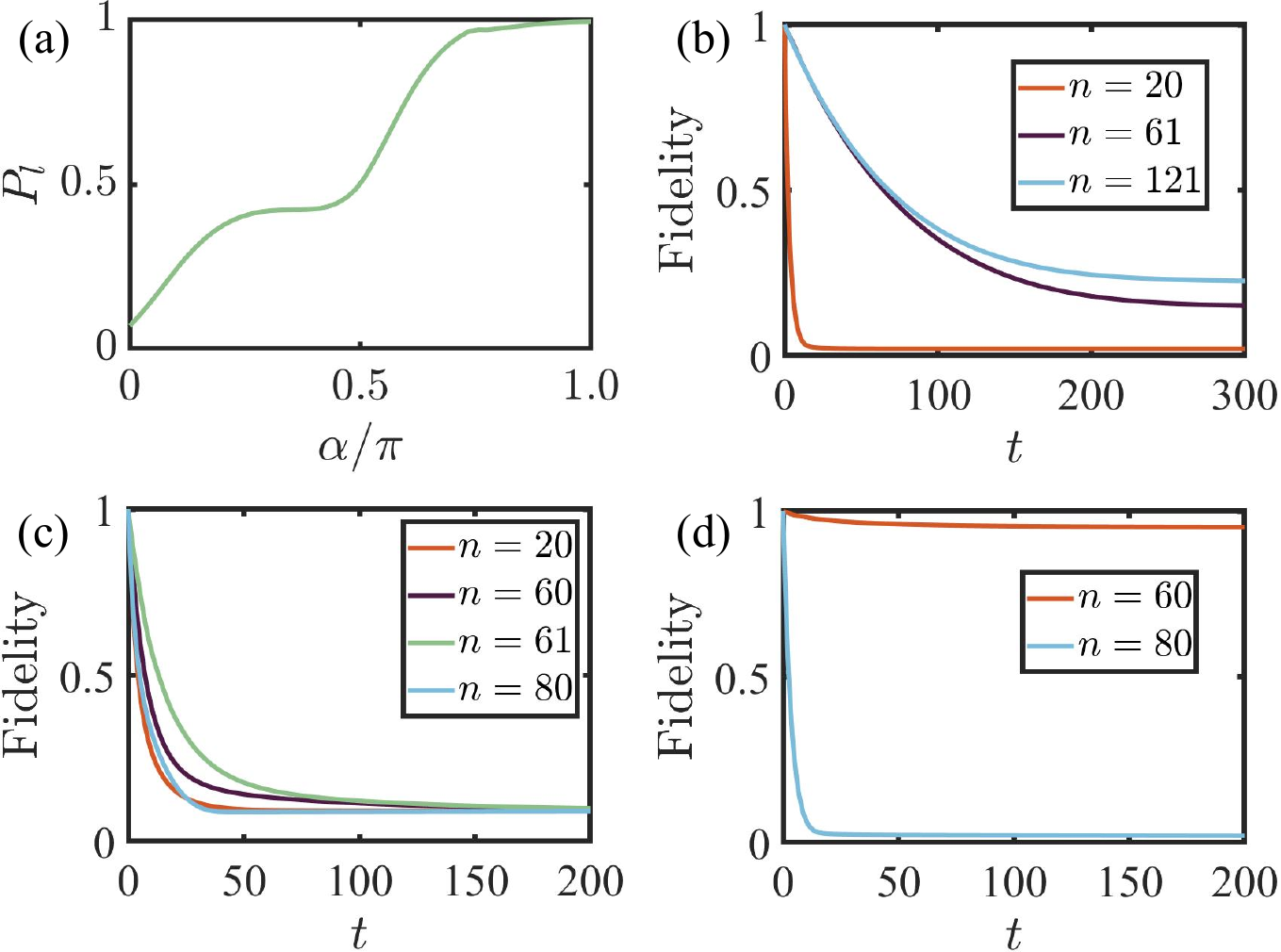}
                \caption{
                    (a) The occupation ration of  localized states $P_l$ in the steady-state density matrix as a function of dissipation $\alpha$. 
                    Time evolution of the quantum fidelity defined by Equation (17), where the initial state is an eigenstate of the Hamiltonian with its index shown in the inset, and the phase: 
                    (b) $\alpha=0$, (c) $\alpha=\pi/2$, (d) $\alpha=\pi$.}
                \label{fig5}
            \end{figure}

When the dissipation phase \( \alpha = 0 \), as shown in the Fig.~\ref{fig5}(b), we choose a localized eigenstate and two extended eigenstates as initial states respectively, with the former in the low-energy localized region ($n=20$) and the latter in middle ($n=61$) and high-energy extended region ($n=121$). One can see the fidelity is keeps high value for extended initial states, while the fidelity rapidly decays to nearly zero if the system is prepared to a localized eigenstate. This is consistent with the fact that localized states occupy a small weight in the steady-state density matrix as shown in the Fig.~\ref{fig5}(a). By contrast, for the case \( \alpha = \pi \), the fidelity is particularly higher for the initial localized state than those extended eigenstates because the steady state is almost entirely composed localized states as plotted in the Fig.~\ref{fig5}(a). 
It should be pointed that above consequences are non-trivial because the localization is fragile under a generic dissipation. For example, if the system couples a bath with local dissipation $O_{j}=n_j$ which describes dephasing at lattice site $j$ or the dissipation phase is chosen by $\alpha=\pi/2$, the system will relax to a maximally mixed state  whose distribution is completely uniform without any localization structure. In this case, the dissipation will erase any initial state information manifesting the value of fidelity is completely independent
of initial states as shown in the Fig.~\ref{fig5}(c).

\begin{figure}[!ht]
        \centering
        \includegraphics[width=1\linewidth]{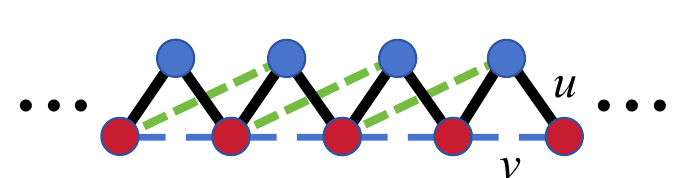}
        \caption{Schematic diagram of dissipation. The green dashed lines indicate position where the dissipation operators load.}
        \label{fig6}
    \end{figure}

Next, we discuss the case of dissipation operators with \( l = 3 \), and the schematic is depicted in the Fig.~\ref{fig6} where the dissipation couples $A$ and $B$ sites in two nearest neighbor unit cells. Then we perform the same calculations as in the case of $l=1$. 

\begin{figure}[!ht]
	\centering
	\includegraphics[width=1\linewidth]{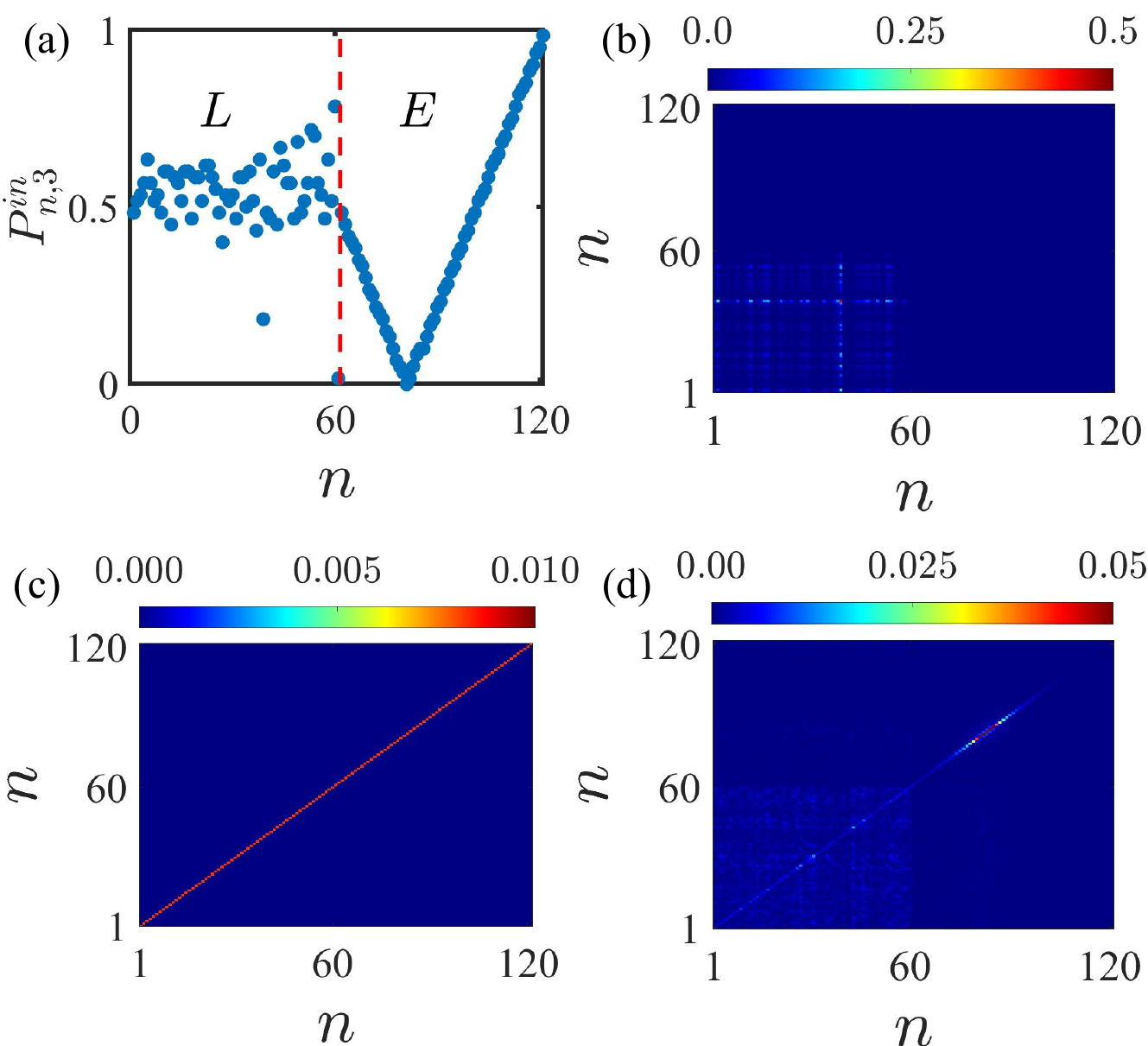}
	\caption{ Steady state induced by dissipation. (a) The proportion $P^{in}_{n,3}$ of in-phase lattice point pairs among all dissipation-applied lattice point pairs in each eigenstate. The localized states and extended states are located on the left and right sides of the red dashed line, respectively. Steady-state density matrix corresponding to the dissipation phase (b) $\alpha=0$, (c) $\alpha=\pi/2$, (d) $\alpha=\pi$.}
	\label{fig7}
\end{figure}

\begin{figure}[!h]
	\centering
	\includegraphics[width=1\linewidth]{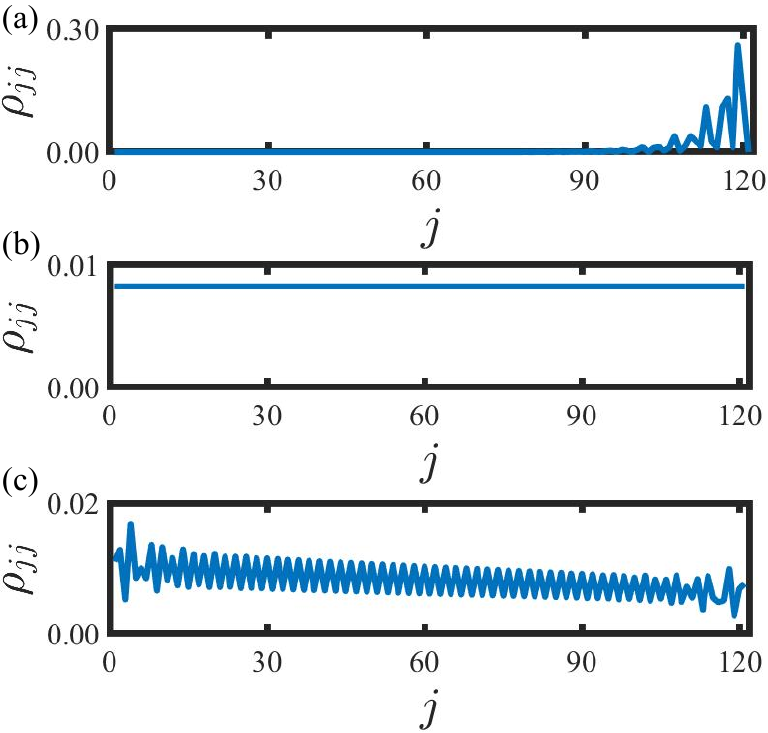}
	\caption{Distribution of the diagonal elements of the density matrix in the lattice basis. The distribution corresponding to (a) $\alpha=0$, (b) $\alpha=\pi/2$, (c) $\alpha=\pi$.}
	\label{fig8}
\end{figure}

When the dissipation phase $\alpha$ is set to $0$, the steady-state density matrix of the system predominantly occupies eigenstates with lower energies as shown in Fig.~\ref{fig7}(b). In contrast, when the dissipation phase equals $\pi$ (Fig.~\ref{fig7}(b)), the steady-state density matrix is primarily composed of a few high-energy eigenstates. This distribution difference can be explained by the in-phase pairs configuration shown in Fig.~\ref{fig7}(a). Specifically, at zero dissipation phase, the dissipation drives the system from out-of-phase pairs toward states dominated by in-phase pairs. Since low-energy eigenstates generally exhibit higher proportions of in-phase pairs on the whole, the steady-state consequently consists mainly of localized low-energy states. When the dissipation phase equals $\pi$, the dissipation instead drives the system from in-phase to out-of-phase configurations which results in a steady state with reduced in-phase pair proportions, corresponding to states near the minimum $P_{n,3}^{in}$ values (the trough of V-shape) in Fig.~\ref{fig7}(a). For the case of $\alpha=\pi/2$, the steady-state density matrix is also completely uniform which is same as the case $l=1$ due to the hermiticity of dissipation operators. The localization property of steady-state can be visualized in the real space. We see there is a localized space distribution for $\alpha=0$ (Fig.~\ref{fig8}(a)), and an extended profile for $\alpha=\pi$ as presented in the Fig.~\ref{fig8}(c). A constant distribution occurs when $\alpha=\pi/2$. These spatial distributions  match up with consequences in the distributions on eigenbasis in the Fig.~\ref{fig7}. 
Hence, when the dissipation phase $\alpha$ varies from $0$ to $\pi$, the main components of the steady-state shift from localized states to  extended states which indicates a localization-delocalization transition can be realized by tuning the dissipative phase. 


\begin{figure}[!h]
	\centering
	\includegraphics[width=1\linewidth]{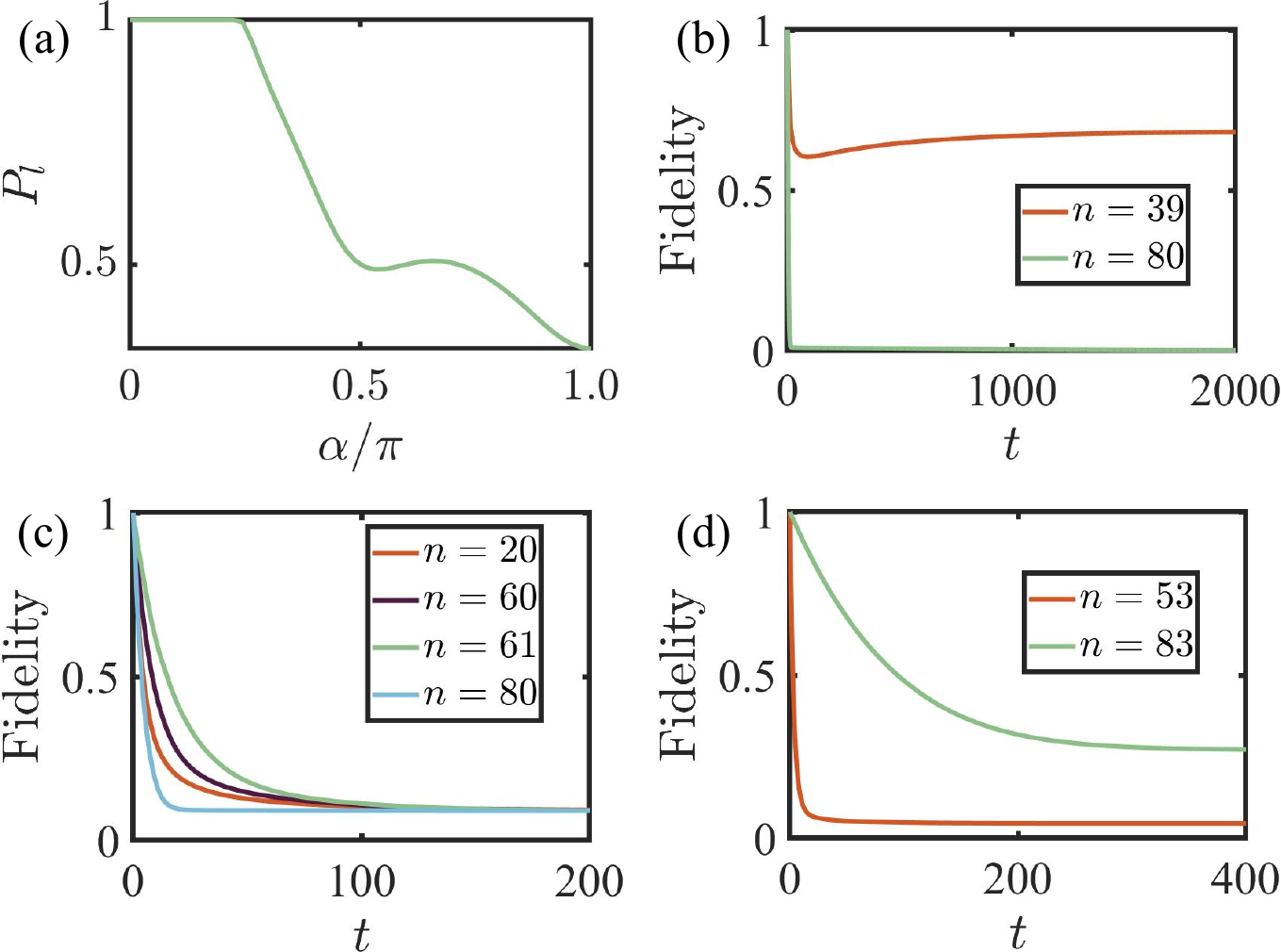}
	\caption{ (a) The sum of the diagonal elements of the density matrix corresponding to the localized states, $P_l$, as a function of $\alpha$. The time evolution of the fidelity defined by Equation (25), where the initial state is an eigenstate of the Tasaki Hamiltonian, with its index shown in the inset, and the phase (b) $\alpha=0$, (c) $\alpha=\pi/2$, (d) $\alpha=\pi$.}
	\label{fig9}
\end{figure}

Dissipation induced localization transition can also be explained from perspective of dynamics.  We calculate time evolution of the quantum fidelity for different dissipation phases and the ratio of localized modes $P_l$ in the Fig.~\ref{fig9}. When the dissipation phase \(\alpha = 0\), as shown in Fig.~\ref{fig9}(b), we choose a localized eigenstate and an extend eigenstates as initial states. It can be found that the quantum fidelity remains high for localized initial state but rapidly decays to zero for the extended eigenstate. This aligns with the consequence that the ratio of localized states $P_l$ is almost equal to $1$, as seen in Fig.~\ref{fig9}(a). In contrast, for \(\alpha = \pi\), the fidelity is significantly higher for the extended initial state than for localized eigenstate (Fig.~\ref{fig9}(c)), because the steady state is predominantly composed of extended states $P_l\ll1$ as shown in the Fig.~\ref{fig9}(a). For the case of $\alpha=pi/2$, we see the fidelity relaxes to identical value for all initial states which is because the steady-state density matrix is proportional to identity matrix.


\par

\section{Conclusion} 
\label{summarize}
In this work, we have demonstrated that tailored dissipative operators can induce a controlled transition between localized and extended phases in a flat-band Tasaki lattice. By tuning the dissipation parameter, the system's steady-state can be selectively driven toward either localized or delocalized eigenstates, independent of initial conditions. We have identified that this transition stems from the phase-sensitive nature of the dissipative operators, which preferentially preserve eigenstates with specific relative phase patterns between coupled sites. The proportion of in-phase site pairs $P^{in}_{n,l}$ serves as an effective indicator for predicting the steady-state configuration. For $l=1$ ($l=3$) and  $\alpha = 0$ ($\alpha=\pi$), the system evolves toward extended states in the high-energy region, while at $\alpha = \pi$ ($\alpha=\pi$), it favors localized states in the low-energy region. The special case of $\alpha = \pi/2$ leads to complete thermalization with a uniform density matrix. 

We also discuss the dissipation-induced transition from quantun dynamics. The quantum fidelity dynamics provide clear experimental signatures of the transition, showing distinct temporal behaviors for different dissipation phases and initial states. Recent progress in quantum experiments has shown that engineered dissipation can be implemented in cold-atom platforms. Optical lattice experiments, for example, have achieved precise manipulation of quantum states via tailored dissipative processes \cite{simulation1,simulation2}. Specifically, spin-dependent potentials and auxiliary lattice sites \cite{Jenkins2022,Chen2022} enable dissipative coupling across neighboring ($l=1$) and next-nearest ($l>1$) sites through optimized laser arrangements \cite{WYC_PRL}. These results establish a promising pathway for experimentally realizing our dissipation-based protocol and detecting emergent localization-delocalization transitions in quantum matter.

Furthermore, the phase-modulated dissipation operators introduced here suggest new opportunities for exploring alternative experimental schemes to prepare targeted quantum states. From a fundamental perspective, our results provide crucial insights into how engineered dissipation affects localization phenomena in flat-band systems, shedding new light on the preservation of quantum coherence in open quantum systems. From a practical standpoint, the demonstrated ability to precisely control the system's asymptotic state through dissipation parameters establishes a powerful toolkit for quantum engineering applications. 
These advances significantly deepen our understanding of controlled quantum dynamics in dissipative environments while simultaneously opening novel pathways for quantum information processing and engineered quantum materials with tunable transport properties.
%
%
%
\\

\section*{Acknowledgements}
The work is supported by the National Key R\&D Program of China under Grant No.2022YFA1405800 and National Natural Science Foundation of China (Grant No. 12304290). LP also acknowledges support from the Fundamental Research Funds for the Central Universities. \\

\textit{Note added.} During the final stages of preparing this manuscript, we became aware that Ref.~\cite{Flat_Wang} has also investigated dissipation-induced localization phenomena in flat-band systems.

\appendix
\section{Derivation of the Liouvillian superoperator } \label{App1}

In this appendix, we give a derivation of the Liouvillian superoperator formula \ref{Liouvillian} in the main text. We start from the Lindblad master equation 
\begin{align}
	\frac{d\rho(t)}{dt} = -i[H_S, \rho(t)] + \sum_{j} \sum_{m=1}^M \Gamma_j^{(m)} &\Big( O_j^{(m)} \rho O_j^{(m)\dagger}\nonumber \\
	&- \frac{1}{2} \{ O_j^{(m)\dagger} O_j^{(m)}, \rho \} \Big).\nonumber 
\end{align}
and then utilizing the Choi-Jamiołkowski isomorphism, the density matrix \(\rho\) can be written as vectorized form \(|\rho\rangle = \mathrm{vec}(\rho)\), where \(\mathrm{vec}(\rho)\) transforms columns of $\rho$ to a single column vector in which way each item in the Lindblad equation becomes 

\begin{align} \mathrm{vec}(A \rho B) &= (B^{\mathrm{T}} \otimes A) \, \mathrm{vec}(\rho), \nonumber\\ \mathrm{vec}([A, \rho]) &= (A \otimes I - I \otimes A^{\mathrm{T}}) \, \mathrm{vec}(\rho), \nonumber\\ \mathrm{vec}(\{A, \rho\}) &= (A \otimes I + I \otimes A^{\mathrm{T}}) \, \mathrm{vec}(\rho). 
\end{align}

Hence, the unitary part  \(-i[H_S, \rho]\) becomes
\begin{align}
	\mathrm{vec}(-i[H_S, \rho]) = -i(H_S \otimes I - I \otimes H_S^{\mathrm{T}}) \, |\rho\rangle,
\end{align}
and the dissipative part is transformed as
\begin{align}
	\mathrm{vec}(O_j^{(m)} \rho O_j^{(m)\dagger}) &= (O_j^{(m)} \otimes O_j^{(m)*}) \, |\rho\rangle, \nonumber \\
	\mathrm{vec}(\{O_j^{(m)\dagger} O_j^{(m)}, \rho\}) &= (O_j^{(m)\dagger} O_j^{(m)} \otimes I + I \otimes (O_j^{(m)\dagger} O_j^{(m)})^{\mathrm{T}}) \, |\rho\rangle.
\end{align}

Therefore, the Lindblad equation can be rewritten by means of the Liouvillian superoperator and the vectorized density matrix $ \frac{d|\rho\rangle}{dt} = \mathscr{L} |\rho\rangle$, where the total Liouvillian superoperator $\mathscr{L}$ includes the unitary and dissipative parts 
\begin{widetext}
	\begin{align}
		\mathscr{L} = -i(H_S \otimes I - I \otimes H_S^{\mathrm{T}}) + \sum_{j} \sum_{m=1}^M\Gamma_j^{(m)} \left[ 2 O_j^{(m)} \otimes O_j^{(m)*}- O_j^{(m)\dagger} O_j^{(m)} \otimes I - I \otimes \Big(O_j^{(m)\dagger} O_j^{(m)}\Big)^{\mathrm{T}} \right]. \nonumber
	\end{align} 
\end{widetext}
This is the full Liouvillian superoperator Eq.\eqref{Liouvillian} in the main text.


\begin{thebibliography}{36}
\bibitem{Anderson1958} P. W. Anderson, 
Absence of Diffusion in Certain Random Lattices,
\href{https://journals.aps.org/pr/abstract/10.1103/PhysRev.109.1492}{Phys. Phys. Rev. {\bf 109}, 1492 (1958).}	
\bibitem{Wiersma2009} A. Lagendijk, B. Tiggelen, and D. S. Wiersma, Fifty years of Anderson localization, Phys. Today {\bf 62}, 24 (2009).
 
\bibitem{RMP1} P. A. Lee and T. V. Ramakrishnan, Disordered electronic systems, Rev. Mod. Phys. {\bf 57}, 287 (1985).
\bibitem{RMP2} F. Evers and A. D. Mirlin, Anderson transitions, Rev. Mod. Phys. {\bf 80}, 1355 (2008).

\bibitem{Exp1} 
G. Barontini, R. Labouvie, F. Stubenrauch, A. Vogler, V. Guarrera, and H. Ott,
Controlling the Dynamics of an Open Many-Body Quantum System with Localized Dissipation,
\href{https://journals.aps.org/prl/abstract/10.1103/PhysRevLett.110.035302}{Phys. Rev. Lett. {\bf 110}, 035302 (2013).}

\bibitem{Exp2} 
B. Yan, S. A. Moses, B. Gadway, J. P. Covey, K. R. Hazzard, A. M. Rey, D. S. Jin, and J. Ye,
Observation of dipolar spin-exchange interactions with lattice-confined polar molecules,
\href{https://www.nature.com/articles/nature12483}{Nature (London) {\bf 501}, 521 (2013).}

\bibitem{Exp3} 
Y. S. Patil, S. Chakram, and M. Vengalattore,
Measurement-Induced Localization of an Ultracold Lattice Gas,
\href{https://journals.aps.org/prl/abstract/10.1103/PhysRevLett.115.140402}{Phys. Rev. Lett. {\bf 115}, 140402 (2015).}

\bibitem{Exp4} 
R. Labouvie, B. Santra, S. Heun, and H. Ott,
Bistability in a Driven-Dissipative Superfluid,
\href{https://journals.aps.org/prl/abstract/10.1103/PhysRevLett.116.235302}{Phys. Rev. Lett. {\bf 116}, 235302 (2016).}

\bibitem{Exp5} 
T. Tomita, S. Nakajima, I. Danshita, Y. Takasu, and Y. Takahashi,
Observation of the Mott insulator to superfluid crossover of a driven-dissipative Bose-Hubbard system,
\href{https://www.science.org/doi/full/10.1126/sciadv.1701513}{Sci. Adv. {\bf 3}, e1701513 (2017).}

\bibitem{Exp6} 
H. P. L{\"u}schen, P. Bordia, S. S. Hodgman, M. Schreiber, S. Sarkar, A. J. Daley, M. H. Fischer, E. Altman, I. Bloch, and U. Schneider,
Signatures of Many-Body Localization in a Controlled Open Quantum System,
\href{https://journals.aps.org/prx/abstract/10.1103/PhysRevX.7.011034}{Phys. Rev. X {\bf 7}, 011034 (2017).}

\bibitem{Exp7} 
K. Sponselee, L. Freystatzky, B. Abeln, M. Diem, B. Hundt, A. Kochanke, T. Ponath, B. Santra, L. Mathey, K. Sengstock, and C. Becker,
Dynamics of ultracold quantum gases in the dissipative Fermi–Hubbard model,
\href{https://iopscience.iop.org/article/10.1088/2058-9565/aadccd/meta}{Quantum Sci. Technol. {\bf 4}, 014002 (2018).}

\bibitem{Exp8} 
T. Tomita, S. Nakajima, Y. Takasu, and Y. Takahashi,
Dissipative Bose-Hubbard system with intrinsic two-body loss,
\href{https://journals.aps.org/pra/abstract/10.1103/PhysRevA.99.031601}{Phys. Rev. A {\bf 99}, 031601(R) (2019).}


\bibitem{OpenMB1} K. Kawabata, Y. Ashida, and M. Ueda,
Information Retrieval and Criticality in Parity-Time-Symmetric Systems,
\href{https://journals.aps.org/prl/abstract/10.1103/PhysRevLett.119.190401}{Phys. Rev. Lett. {\bf 119}, 190401 (2017).}

\bibitem{OpenMB2} Y. Ashida, S. Furukawa, and M. Ueda,
Parity-time-symmetric quantum critical phenomena,
\href{https://www.nature.com/articles/ncomms15791}{Nat. Commun. {\bf 8}, 15791 (2017).}

\bibitem{OpenMB3} K. Yamamoto, M. Nakagawa, K. Adachi, K. Takasan, M. Ueda, and N. Kawakami,
Theory of Non-Hermitian Fermionic Superfluidity with a Complex-Valued Interaction,
\href{https://journals.aps.org/prl/abstract/10.1103/PhysRevLett.123.123601}{Phys. Rev. Lett. {\bf 123}, 123601 (2019).}

\bibitem{OpenMB4} N. Okuma and M. Sato,
Topological Phase Transition Driven by Infinitesimal Instability: Majorana Fermions in Non-Hermitian Spintronics,
\href{https://journals.aps.org/prl/abstract/10.1103/PhysRevLett.123.097701}{Phys. Rev. Lett. {\bf 123}, 097701 (2019).}

\bibitem{OpenMB5} L. Zhou and X. Cui,
Enhanced Fermion Pairing and Superfluidity by an Imaginary Magnetic Field,
\href{https://www.sciencedirect.com/science/article/pii/S2589004219300975}{iScience {\bf 14}, 257 (2019).}

\bibitem{OpenMB6} L. Pan, X. Chen, Y. Chen, H. Zhai,
Non-Hermitian linear response theory,
\href{https://www.nature.com/articles/s41567-020-0889-6}{Nat. Phys. {\bf 16}, 767 (2020).}

\bibitem{OpenMB7} Z. Cai and T. Barthel,
Algebraic versus Exponential Decoherence in Dissipative Many-Particle Systems,
\href{https://journals.aps.org/prl/abstract/10.1103/PhysRevLett.111.150403}{Phys. Rev. Lett. {\bf 111}, 150403 (2013).}

\bibitem{OpenMB8} K. Yamamoto, M. Nakagawa, N. Tsuji, M. Ueda, and N. Kawakami,
Collective Excitations and Nonequilibrium Phase Transition in Dissipative Fermionic Superfluids,
\href{https://journals.aps.org/prl/abstract/10.1103/PhysRevLett.127.055301}{Phys. Rev. Lett. {\bf 127}, 055301 (2021).}	



\bibitem{OpenMB9} R. Hanai, A. Edelman, Y. Ohashi, and P. B. Littlewood,
Non-Hermitian Phase Transition from a Polariton Bose-Einstein Condensate to a Photon Laser,
\href{https://journals.aps.org/prl/abstract/10.1103/PhysRevLett.122.185301}{Phys. Rev. Lett. {\bf 122}, 185301 (2019).}


\bibitem{OpenMB10} K. L. Zhang and Z. Song,
Quantum Phase Transition in a Quantum Ising Chain at Nonzero Temperatures,
\href{https://journals.aps.org/prl/abstract/10.1103/PhysRevLett.126.116401}{Phys. Rev. Lett. 126, 116401 (2021).}

\bibitem{OpenMB11} D. Sticlet, B. D{\'o}ra, and C. P. Moca,
Kubo Formula for Non-Hermitian Systems and Tachyon Optical Conductivity,
\href{https://journals.aps.org/prl/abstract/10.1103/PhysRevLett.128.016802}{Phys. Rev. Lett. 128, 016802 (2022).}

\bibitem{OpenMB12} B. D{\'o}ra, and C. P. Moca,
Quantum Quench in $\mathscr{PT}$
-Symmetric Luttinger Liquid,
\href{https://journals.aps.org/prl/abstract/10.1103/PhysRevLett.124.136802}{Phys. Rev. Lett. 124, 136802 (2020).}

\bibitem{OpenMB13} {\'A}. B{\'a}csi, C. P. Moca, and B. D{\'o}ra,
Dissipation-Induced Luttinger Liquid Correlations in a One-Dimensional Fermi Gas,
\href{https://journals.aps.org/prl/abstract/10.1103/PhysRevLett.124.136401}{Phys. Rev. Lett. 124, 136401 (2020).}

\bibitem{OpenMB14} L. Pan, X. Wang, X. Cui, and S. Chen, 
Interaction-induced dynamical $\mathscr{PT}$
-symmetry breaking in dissipative Fermi-Hubbard models,
\href{https://journals.aps.org/pra/abstract/10.1103/PhysRevA.102.023306}{Phys. Rev. A {\bf 102}, 023306 (2020).}

\bibitem{OpenMB15} X. Z. Zhang and Z. Song,
Dynamical preparation of a steady off-diagonal long-range order state in the Hubbard model with a local non-Hermitian impurity,
\href{https://journals.aps.org/prb/abstract/10.1103/PhysRevB.102.174303}{Phys. Rev. B {\bf 102}, 174303 (2020).}

\bibitem{OpenMB16} K. Yang, S. C. Morampudi, and E. J. Bergholtz,
Exceptional Spin Liquids from Couplings to the Environment,
\href{https://journals.aps.org/prl/abstract/10.1103/PhysRevLett.126.077201}{Phys. Rev. Lett. {\bf 126}, 0772012 (2021).}


\bibitem{OpenMB17} M. Nakagawa, N. Kawakami, and M. Ueda,
Exact Liouvillian Spectrum of a One-Dimensional Dissipative Hubbard Model,
\href{https://journals.aps.org/prl/abstract/10.1103/PhysRevLett.126.110404}{Phys. Rev. Lett. 126, 110404 (2021).}

\bibitem{OpenMB18} M. Nakagawa, N. Tsuji, N. Kawakami, and M. Ueda, 
Dynamical Sign Reversal of Magnetic Correlations in Dissipative Hubbard Models,
\href{https://journals.aps.org/prl/abstract/10.1103/PhysRevLett.124.147203}{Phys. Rev. Lett. 124, 147203 (2020).}

\bibitem{OpenMB19} L. S{\'a}, P. Ribeiro, and T. Prosen,
Complex Spacing Ratios: A Signature of Dissipative Quantum Chaos,
\href{https://journals.aps.org/prx/abstract/10.1103/PhysRevX.10.021019}{Phys. Rev. X {\bf 10}, 021019 (2020).}

\bibitem{OpenMB20} J. Li, T. Prosen, and A. Chan,
Spectral Statistics of Non-Hermitian Matrices and Dissipative Quantum Chaos,
\href{https://journals.aps.org/prl/abstract/10.1103/PhysRevLett.127.170602}{Phys. Rev. Lett. {\bf 127}, 170602 (2021).} 

\bibitem{OpenMB21} S. Longhi,
Phase transitions and bunching of correlated particles in a non-Hermitian quasicrystal,
\href{https://journals.aps.org/prb/abstract/10.1103/PhysRevB.108.075121}{Phys. Rev. B {\bf 108}, 075121 (2023).}

\bibitem{OpenMB22} T. Mori,
Liouvillian-gap analysis of open quantum many-body systems in the weak dissipation limit,
\href{https://arxiv.org/abs/2311.10304}{arXiv:2311.10304 (2023).}

\bibitem{OpenMB23} C.-Z. Lu, X. Deng, S.-P. Kou, G. Sun,
Unconventional many-body phase transitions in a non-Hermitian Ising chain,
\href{https://doi.org/10.1103/PhysRevB.110.014441}{Phys. Rev. B \bf{110}, 014441 (2024).}


\bibitem{OpenMB25} Non-Hermitian skin effect in a one-dimensional interacting Bose gas,
L. Mao, Y. Hao, and L. Pan,
\href{https://journals.aps.org/pra/abstract/10.1103/PhysRevA.107.043315}{Phys. Rev. A {\bf 107}, 043315 (2023).}

\bibitem{OpenMB26} L. Mao, X. Yang, M.-J. Tao, H. Hu, and L. Pan, Liouvillian skin effect in a one-dimensional open many-body quantum system with generalized boundary conditions,
\href{https://journals.aps.org/prb/abstract/10.1103/PhysRevB.110.045440}{Phys. Rev. B {\bf 110}, 045440 (2024).}




\bibitem{OpenMBL1} R. Hamazaki, K. Kawabata, and M. Ueda,
Non-Hermitian Many-Body Localization,
\href{https://journals.aps.org/prl/abstract/10.1103/PhysRevLett.123.090603}{Phys. Rev. Lett. {\bf 123}, 090603 (2019).} 

\bibitem{OpenMBL2} L.-J. Zhai, S. Yin, and G.-Y. Huang,
Many-body localization in a non-Hermitian quasiperiodic system,
\href{https://journals.aps.org/prb/abstract/10.1103/PhysRevB.102.064206}{Phys. Rev. B {\bf 102}, 064206 (2020).}


\bibitem{OpenMBL3} K. Suthar, Y.-C. Wang, Y.-P. Huang, H.-H. Jen, and J.-S. You, Non-Hermitian many-body localization with open boundaries,
\href{https://doi.org/10.1103/PhysRevB.106.064208}{Phys. Rev. B {\bf 106}, 0642085 (2022).}

\bibitem{OpenMBL4} C. Ehrhardt and J. Larson, Exploring the impact of fluctuation-induced criticality on non-hermitian skin effect and quantum sensors, \href{https://arxiv.org/abs/2310.18259}{arXiv:2310.18259 (2023).}


\bibitem{OpenMBL5} F. Roccati, F. Balducci, R. Shir, and A. Chenu, Diagnosing non-Hermitian many-body localization and quantum chaos via singular value decomposition,  \href{https://journals.aps.org/prb/abstract/10.1103/PhysRevB.109.L140201}{Phys. Rev. B  {\bf 109}, L140201 (2024).}


\bibitem{OpenDisorder1} D. A. Huse, R. Nandkishore, F. Pietracaprina, V. Ros, and A. Scardicchio, Localized systems coupled to small baths: From Anderson to Zeno, \href{https://journals.aps.org/prb/abstract/10.1103/PhysRevB.92.014203}{Phys. Rev. B {\bf 92}, 014203 (2015).}



\bibitem{OpenDisorder2} A. Purkayastha, A. Dhar, and M. Kulkarni, Nonequilibrium phase diagram of a one-dimensional quasiperiodic system with a single-particle mobility edge,  \href{https://journals.aps.org/prb/abstract/10.1103/PhysRevB.96.180204}{Phys. Rev. B {\bf 96}, 180204(R) (2017).}

\bibitem{OpenDisorder3}  V. Balachandran, S. R. Clark, J. Goold, and D. Poletti, Energy Current Rectification and Mobility Edges, \href{https://journals.aps.org/prl/abstract/10.1103/PhysRevLett.123.020603}{Phys. Rev. Lett. {\bf 123}, 020603 (2019).}

\bibitem{OpenDisorder4} C. Chiaracane, M. T. Mitchison, A. Purkayastha,
G. Haack, and J. Goold, Quasiperiodic quantum heat engines with a mobility edge, \href{https://journals.aps.org/prresearch/abstract/10.1103/PhysRevResearch.2.013093}{Phys. Rev. Research {\bf 2}, 013093 (2020).}

\bibitem{OpenDisorder5} M. Balasubrahmaniyam, S. Mondal, and S. Mujumdar, Necklace-State-Mediated Anomalous Enhancement of Transport in Anderson-Localized non-Hermitian Hybrid Systems, \href{https://journals.aps.org/prl/abstract/10.1103/PhysRevLett.124.123901}{Phys. Rev. Lett. {\bf 124}, 123901 (2020).}

\bibitem{OpenDisorder6} S. Weidemann, M. Kremer, S. Longhi, and A. Szameit, Coexistence of dynamical delocalization and spectral localization through stochastic dissipation, \href{https://www.nature.com/articles/s41566-021-00823-w}{Nat. Photon. {\bf 15}, 576 (2021).}

\bibitem{OpenDisorder7} A. M. Lacerda, J. Goold, and G. T. Landi, Dephasing enhanced transport in boundary-driven quasiperiodic chains, \href{https://journals.aps.org/prb/abstract/10.1103/PhysRevB.104.174203}{Phys. Rev. B {\bf 104}, 174203 (2021).}

\bibitem{OpenDisorder8} D. Dwiputra and F. P. Zen, Environment-assisted quantum transport and mobility edges, \href{https://journals.aps.org/pra/abstract/10.1103/PhysRevA.104.022205}{Phys. Rev. A {\bf 104}, 022205 (2021).}


\bibitem{OpenDisorder9} M. Saha, B. P. Venkatesh, and B. K. Agarwalla, Quantum transport in quasiperiodic lattice systems in the presence of B\"{u}ttiker probes, \href{https://journals.aps.org/prb/abstract/10.1103/PhysRevB.105.224204}{Phys. Rev. B {\bf 105}, 224204 (2022).}

\bibitem{OpenDisorder10} C. Chiaracane, A. Purkayastha, M. T. Mitchison, and J. Goold, Dephasing-enhanced performance in quasiperiodic thermal machines,  \href{https://journals.aps.org/prb/abstract/10.1103/PhysRevB.105.134203}{Phys. Rev. B {\bf 105}, 134203 (2022).}

\bibitem{OpenDisorder11} S. Longhi, Anderson Localization in Dissipative Lattices, \href{https://doi.org/10.1002/andp.202200658}{Ann. Phys. {\bf 535}, 2200658 (2023).}

\bibitem{OpenDisorder12} S. Longhi, Dephasing-Induced Mobility Edges in Quasicrystals, 
\href{https://journals.aps.org/prl/abstract/10.1103/PhysRevLett.132.236301}{Phys. Rev. Lett. 132, 236301 (2024).}

\bibitem{OpenDisorder13} X.-P. Jiang, X. Yang, Y. Hu, L. Pan, Dissipation induced ergodic-nonergodic transitions in finite-height mosaic Wannier-Stark lattices, \href{https://arxiv.org/abs/2407.17301}{ arXiv:2407.17301 (2024). }

\bibitem{OpenDisorder14} C. Wang, and X. R. Wang, Anderson localization transitions in disordered non-Hermitian systems with exceptional points \href{https://journals.aps.org/prb/abstract/10.1103/PhysRevB.107.024202}{Phys. Rev. B \textbf{107}, 024202 (2023).}



\bibitem{OpenDisorder15} L.-J. Zhai, S. Yin, and G.-Y. Huang, Many-body localization in a non-Hermitian quasiperiodic system, \href{https://journals.aps.org/prb/abstract/10.1103/PhysRevB.102.064206}{Phys. Rev. B \textbf{102}, 064206 (2020).}

\bibitem{OpenDisorder16} Y.-C. Wang, K. Suthar, H. H. Jen, Y.-T. Hsu, and J.-S. You, Non-Hermitian skin effects on thermal and many-body localized phases, \href{https://journals.aps.org/prb/abstract/10.1103/PhysRevB.107.L220205}{Phys. Rev. B \textbf{107}, L220205 (2023).}

\bibitem{OpenDisorder17} Y. Huang and B. I. Shklovskii, Spectral rigidity of non-Hermitian symmetric random matrices near the Anderson transition, \href{https://journals.aps.org/prb/abstract/10.1103/PhysRevB.102.064212}{Phys. Rev. B \textbf{102}, 064212 (2020).}

\bibitem{OpenDisorder18} Y. Huang and B. I. Shklovskii, Anderson transition in three-dimensional systems with non-Hermitian disorder, \href{https://journals.aps.org/prb/abstract/10.1103/PhysRevB.101.014204}{Phys. Rev. B \textbf{101}, 014204 (2020).} 

\bibitem{OpenDisorder19} Y. Peng, C. Yang, and Y. Wang, Manipulating the relaxation time of boundary-dissipative systems through bond dissipation, \href{https://doi.org/10.1103/PhysRevB.110.104305}{Phys. Rev. B \textbf{110}, 104305 (2024).}

\bibitem{Yusipov17} I. Yusipov, T. Laptyeva, S. Denisov, and M. Ivanchenko, Localization in Open Quantum Systems, \href{https://journals.aps.org/prl/abstract/10.1103/PhysRevLett.118.070402}{Phys. Rev. Lett. {\bf 118}, 070402 (2017).}

 \bibitem{WYC_PRL} Y. Liu, Z. Wang, C. Yang, J. Jie, and Y. Wang, Dissipation-Induced Extended-Localized Transition, \href{https://journals.aps.org/prl/abstract/10.1103/PhysRevLett.132.216301}{Phys. Rev. Lett. {\bf 132}, 216301 (2024).}
 
 \bibitem{Jiang_3D} X. Yang, X.-P. Jiang, Z. Wei, Y. Wang, and L. Pan, Dissipation induced transition between extension and localization in the three-dimensional Anderson model, \href{https://arxiv.org/abs/2409.20319}{arXiv:2409.20319 (2024).}
 
 \bibitem{Yusipov18} I. Vakulchyk, I. Yusipov, M. Ivanchenko, S. Flach, and S. Denisov, Signatures of many-body localization in steady states of open quantum systems, \href{https://journals.aps.org/prb/abstract/10.1103/PhysRevB.98.020202}{Phys. Rev. B {\bf 98}, 020202(R) (2018).}
 
 \bibitem{WYC_MBL} Y. Hu, C. Yang, and Y. Wang, Can dissipation induce a transition between many-body localized and thermal states? 
 \href{https://arxiv.org/abs/2407.13655}{arXiv:2407.13655 (2024).}
 
 \bibitem{Diss_scar} X.-P. Jiang, M. Xu, X. Yang, H. Hou, Y. Wang, and L. Pan, Robustness of quantum many-body scars in the presence of Markovian bath, \href{https://arxiv.org/abs/2501.00886}{ 	arXiv:2501.00886 (2025).}

\bibitem{BoYan_flatband} Hang Li, Zhaoli Dong, Stefano Longhi, Qian Liang, Dizhou Xie, and Bo Yan, Aharonov-Bohm Caging and Inverse Anderson Transition in Ultracold Atoms,\href{https://doi.org/10.1103/PhysRevLett.129.220403}{Phys. Rev. A {\bf 129}, 220403 (2022).}

 \bibitem{Flatband_exp} C. Zeng, Y.-R. Shi, Y.-Y. Mao, F.-F. Wu, Y.- J. Xie, T. Yuan, W. Zhang, H.-N. Dai, Y.-A. Chen, and J.-W. Pan, W. Zhang, J. W. Pan, Transition from Flat-Band Localization to Anderson Localization in a One-Dimensional Tasaki Lattice, \href{https://doi.org/10.1103/PhysRevLett.132.063401}{Phys. Rev. Lett. {\bf 132}, 063401 (2024).}

\bibitem{Moy1999}
G. M. Moy, J. J. Hope, and C. M. Savage,
Born and Markov approximations for atom lasers,
\href{https://doi.org/10.1103/PhysRevA.59.667}{Phys. Rev. A \textbf{59}, 667 (1999).}

\bibitem{Breuer2002}
H.-P. Breuer and F. Petruccione,
\textit{The Theory of Open Quantum Systems},
(Oxford University Press, Oxford, 2002).

\bibitem{Lindblad1} G. Lindblad, On the generators of quantum dynamical semigroups,
\href{https://link.springer.com/article/10.1007/BF01608499}{Commun. Math. Phys. {\bf 119}, 48 (1976).}

\bibitem{Lindblad2} V. Gorini, A. Kossakowski, and E. C. Sudarsahan,
Completely positive dynamical semigroups of N‐level systems,
\href{https://pubs.aip.org/aip/jmp/article/17/5/821/225427/Completely-positive-dynamical-semigroups-of-N}{J. Math. Phys. {\bf 17}, 821 (1976).}

\bibitem{CJ1} M.-D. Choi,
Completely positive linear maps on complex matrices,
\href{https://www.sciencedirect.com/science/article/pii/0024379575900750?via%3Dihub}{Lin. Alg. Appl. {\bf 10}, 285 (1975).}

\bibitem{CJ2} A. Jamiołkowski, Linear transformations which preserve trace and positive semidefiniteness of operators,
\href{https://www.sciencedirect.com/science/article/abs/pii/0034487772900110?via%3Dihub}{Rep. Math. Phys. {\bf 3}, 275 (1972).}

\bibitem{Jump1} S. Diehl, A. Micheli, A. Kantian, B. Kraus, H. P. Büchler, and P. Zoller, Quantum states and phases in driven open quantum systems with cold atoms,  \href{https://www.nature.com/articles/nphys1073}{Nat. Phys. {\bf 4}, 878 (2008).}

\bibitem{Jump2} B. Kraus, H. P. Büchler, S. Diehl, A. Kantian, A. Micheli, and P. Zoller, Preparation of entangled states by quantum Markov processes,  \href{https://journals.aps.org/pra/abstract/10.1103/PhysRevA.78.042307}{Phys. Rev. A {\bf 78}, 042307 (2008).}

\bibitem{BHchain} D. Marcos, A. Tomadin, S. Diehl, and P. Rabl,
Photon condensation in circuit quantum electrodynamics by engineered dissipation, \href{https://iopscience.iop.org/article/10.1088/1367-2630/14/5/055005}{New J. Phys. {\bf 14}, 055005 (2012).}

\bibitem{Longhi2023} S. Longhi,  Anderson Localization in Dissipative Lattices, \href{https://doi.org/10.1002/andp.202200658}{Ann. Phys. (Berlin) \textbf{535}, 2200658 (2023).}

\bibitem{Fidelity1}  M. Nielsen and I. Chuang, Quantum Computation
and Quantum Information (Cambridge University Press,
Cambridge, England, 2000).
\bibitem{Fidelity2} P. Zanardi, H. T. Quan, Xiaoguang Wang, and C. P. Sun, Mixed-state fidelity and quantum criticality at finite temperature,
\href{https://doi.org/10.1103/PhysRevA.75.032109}{Phys. Rev. A {\bf 75}, 032109 (2007).}





\bibitem{simulation1}
M. M{\"u}ller, S. Diehl, G. Pupillo, and P. Zoller, 
Engineered Open Systems and Quantum Simulations with Atoms and Ions,
\href{https://arxiv.org/abs/1203.6595}{Adv. At. Mol. Opt. Phys. \textbf{61}, 1-80 (2012).}

\bibitem{simulation2} P. M. Harrington, E. J. Mueller, and K. W. Murch, Engineered dissipation for quantum information science, \href{https://doi.org/10.1038/s42254-022-00494-8}{Nat. Rev.
	Phys. {\bf 4}, 660 (2022).}
 

 


\bibitem{Jenkins2022} 
A. Jenkins, J. W. Lis, A. Senoo, W. F. McGrew, and A. M. Kaufman, 
Ytterbium nuclear-spin qubits in an optical tweezer array, 
\href{https://journals.aps.org/prx/abstract/10.1103/PhysRevX.12.021027}{Phys. Rev. X
	\textbf{12}, 021027 (2022).} 

\bibitem{Chen2022} 
N. Chen, L. Li, W. Huie, M. Zhao, I. Vetter, C. H. Greene, and J. P. Covey, 
Analyzing the Rydberg-based optical-metastable-ground architecture for \(^{171}\)Yb nuclear spins, 
\href{https://journals.aps.org/pra/abstract/10.1103/PhysRevA.105.052438}{Phys. Rev. A
	\textbf{105}, 052438 (2022).} 

\bibitem{Flat_Wang} Y. Hu, C. Yang, and Y. Wang, Dissipation-Driven Transition of Particles from Dispersive to Flat Bands, \href{https://arxiv.org/abs/2504.00796}{arXiv:2504.00796}

\end{thebibliography}
\end{document}